\documentclass[12pt]{article}

\newfont{\bit}{cmbxti10 scaled 1728}

\renewcommand{\thefootnote}{\fnsymbol{footnote}}
\begin{document}
\newpage
\pagestyle{empty}

\begin{center}
{\LARGE {Generalized Symmetries\\
 of \\
Impulsive Gravitational Waves\\ 
}}

\vspace{1cm}
{\large
 Peter C. AICHELBURG 
 \footnote[1]{ e-mail: pcaich @ pap.univie.ac.at}
  \footnote[4]{supported in part by the FUNDACION FEDERICO}
}\\
{\em 
 Institut f\"ur Theoretische Physik, Universit\"at Wien\\
 Boltzmanngasse 5, A - 1090 Wien, AUSTRIA 
 }\\[.5cm]
{\em and}\\[.5cm]
{\large Herbert BALASIN
\footnote[2]{e-mail: hbalasin @ email.tuwien.ac.at}
\footnote[3]{supported by the APART-program of the 
Austrian Academy of Sciences}
}\\ 
{\em
 Institut f\"ur Theoretische Physik, Technische Universit\"at Wien\\
 Wiedner Hauptstra{\ss}e 8--10, A - 1040 Wien, AUSTRIA 
}\\[.8cm]
{{\bf Dedication:} We feel honored to dedicate this article to \\
Andrzej Trautman on the occasion of his $8^2$-th birthday}
\end{center}
\begin{abstract}
We generalize previous \cite{AiBa2} work on the classification of 
($C^\infty$) symmetries of plane-fronted waves with an impulsive profile.
Due to the specific form of the profile it is possible to extend
the group of normal-form-preserving diffeomorphisms to include 
non-smooth transformations. This extension
entails a richer structure of the symmetry algebra 
generated by the (non-smooth) Killing vectors.

\end{abstract}

\rightline{UWThPh -- 1996 -- 28}
\rightline{TUW 96 -- 07}
\rightline{April 1996}

\renewcommand{\thefootnote}{\arabic{footnote}}
\setcounter{footnote}{0}
\newpage  
\pagebreak
\pagenumbering{arabic}
\pagestyle{plain}
\renewcommand{\baselinestretch}{1}
\small\normalsize 
\section*{\Large\bit Introduction}

In a previous paper \cite{AiBa2} the authors considered symmetries of 
impulsive plane-fronted waves with parallel rays. The motivation for this
study was the failure of the classical work by Jordan, Ehlers and Kundt 
(JEK) \cite{JEK} for wave profiles which are generalized functions 
\cite{GeSh}. 
In order to solve this problem a new systematic approach for the 
classification of general pp-waves was presented \cite{AiBa2}, which was 
in the following applied to the simplest class of impulsive spacetimes, 
where the wave is concentrated on a single null hyperplane. It 
was shown that even vacuum spacetimes within this class contain new symmetries 
which have no non-impulsive analogue.\par
In the above mentioned  analysis it was assumed that the symmetry generating 
vector fields are smooth, because only then it is guaranteed that in general 
Killing's equation is well-defined. In present paper we relax this 
condition, i.e.~we do not require $C^\infty$-Killing fields from the outset, 
but rather adapt to the profile of the wave under consideration. What we 
require is that for a given profile, each term 
in Killing's equation is well defined in the sense of generalized functions.
Again we apply this reasoning to our prototype of impulsive waves. The 
analysis shows that in this class there exist, besides of the smooth Killing 
fields found in \cite{AiBa2}, symmetries that are only continuous or even 
discontinuous. We give a complete classification for these spacetimes in 
terms of their symmetries and show that the structure of these ``generalized 
symmetries'' is much richer. As in \cite{AiBa2} we do not impose Einsteins 
vacuum equation.\par
It was argued by R.~Penrose \cite{Pen}  that although the impulsive waves 
have a delta-like metric component in adapted coordinates, the metric can 
be made continuous by going over to the so-called Rosen form \cite{Rosen}.
This of course involves a discontinuous transformation by which the 
originally smooth Killing fields become non-smooth. We show this explicitly 
for impulsive plane waves, which admit a five parameter group 
of isometries.\par
The paper is organized as follows: in section 1 we introduce pp-wave 
spacetimes in adapted  (normal form) coordinates. We generalize the so-called 
normal-form-preserving diffeomorphisms to non-smooth ones for waves with  
impulsive profile.  In section 2 we investigate the abstract group 
structure of these transformations and classify its Lie-algebra in terms of 
adjoint orbits. Then  we impose the ``canonical'' representative of 
each orbit as generator of an isometry and obtain the corresponding 
wave profile. Since, as pointed out in \cite{AiBa2},  any impulsive wave 
admits a three parameter group of isometries,  this method leads to waves with
at least four Killing vectors. In section 3 spacetimes belonging to even 
higher symmetry classes are obtained from the combination of different 
generators. 
This is a non-trivial procedure since in general the generators cannot be 
transformed to their canonical form simultaneously. 
The complete classification is presented in a table at 
the end of this section. Finally, in section 4  we transform the plane 
impulsive wave to the Rosen form.  

\section*{\Large\bit 1) Impulsive pp-waves}

In this section we briefly recall the definition of pp-waves, especially
impulsive ones. With respect to adapted coordinates the metric takes 
the ``normal'' form \cite{Brink}
\begin{equation}\label{NF}
ds^2=-dudv  + dx^2 + f(u,x) du^2,
\end{equation}
where $x$ are flat coordinates in the two-dimensional spacelike subspace
$u=v=const$.
Therefore a pp-wave is completely characterized by its profile-function 
$f(u,x)$, which is unique up to normal form (nfp) preserving 
diffeomorphisms \cite{JEK, AiBa2}. 
The latter are given by
\begin{eqnarray}
\tilde{u}&=&au+b,\nonumber\\
\tilde{v}&=&\frac{1}{a}\left(v + 2(\Omega x\cdot d'(u) + n(u)) \right),
        \nonumber\\
\tilde{x}&=&\Omega x + d(u).
\end{eqnarray}
Here we follow the notation introduced in \cite{AiBa2} where $a$ and $b$
are constants, $\Omega$ is in $SO(2)$ and $d(u)$ and $n(u)$ denote a 
two-dimensional vector field and a function respectively.
Impulsive pp-waves are further restricted by requiring their curvature 
to be concentrated on the null hyperplane $u=0$. 
This implies that the profile is of the form
\begin{equation}\label{DisPro}
  f(u,x)=f(x)\delta (u),
\end{equation}
where $f(x)$ will in the following , by a slight abuse of language,
also be referred to as the profile.\par
For a profile that is genuinely distributional 
in nature only $C^\infty$ nfp-diffeomorphisms are allowed.
However, taking into account that $\delta(u)$ is an order zero distribution
it is possible  to extend the class of nfp-transformations by including
terms which are non-smooth at $u=0$.
Let us therefore examine the transformation behavior of the profile
with respect to a nfp-transformation 
\begin{eqnarray}\label{TraPro}
  \tilde{f}(u,x) &=& a^2 \delta(au) f(\Omega x + d(u)) + d'^2(u)
  -2(d''(u)\cdot\Omega x + n'(u))\nonumber\\
  &=&a\delta(u)f(\Omega x + d(0)) + d'^2(u) - 2(d''(u) \cdot\Omega x + n'(u)),
\end{eqnarray}
where we have set the parameter $b=0$ since its sole effect amounts in a
shift of the location of the pulse.
In order to produce at most delta-function terms the non-smooth
generalization has to be of the form
\newline\noindent 
\begin{minipage}[c]{6cm}
\begin{eqnarray*}
  d(u)&=& d_s(u) + \theta(u) d_+(u),\nonumber\\
  n(u)&=& n_s(u) + \theta(u) n_+(u),
\end{eqnarray*}
\end{minipage} 
\begin{minipage}[c]{7cm}
  \begin{displaymath}
    d_s,d_+,n_s,n_+ \in C^\infty,
  \end{displaymath}
\end{minipage}
\newline
\vspace{0.1cm}

\noindent
where $\theta(u)$ is the Heaviside step function.
The second line of (\ref{TraPro}) tells us that $d(u)$ has to be 
well defined at $u=0$ thus requiring $d_+(0)=0$. As a consequence 
the transformed profile becomes
\begin{eqnarray}\label{trapro1}
  \tilde{f}(u,x)&=&\delta(u)(af(\Omega x + d_0) - 
  2(d_1^+\cdot\Omega x + n_0^+))  \nonumber\\
  &+&\theta(u)({d_+'}^2(u) + 2d_s'(u)\cdot d_+'(u) -2(d_+''(u)\cdot \Omega x 
  + n_+'(u)))\nonumber\\
  &+& ({d_s'}^2(u) -2(d_s''(u)\cdot\Omega x + n_s'(u))),
\end{eqnarray}
where we adopted the following notation: indices $0$ and $1$ denote the
value of the function and its derivative at $u=0$ respectively. 
>From $\tilde{f}(u,x)=\tilde{f}(x)\delta(u)$ and
since the smooth part of (\ref{trapro1}) has to vanish separately, we find 
\begin{eqnarray}\label{Smoo}
  d_s''(u)=0 &\Longrightarrow& d_s=d_0+ud_1\nonumber\\
  n_s'(u)=\frac{1}{2}{d_s'}^2(u) &\Longrightarrow& n_s(u)=n_0+ 
  \frac{u}{2} d_1^2.
\end{eqnarray}
The remaining terms may be split into those contributing for
positive values of $u$ and those which are concentrated at $u=0$.
\begin{eqnarray}\label{Pos}
  d_+''(u)=0 &\Longrightarrow& d_+(u)= d_1^+ u\nonumber\\
  n_+'(u)=d_1\cdot d_1^+ + \frac{1}{2}{d_1^+}^2 &\Longrightarrow&n_+(u)=
  n_0^+ + \frac{u}{2}({d_1^+}^2 + 2 d_1\cdot d_1^+).
\end{eqnarray}
Finally, we obtain from (\ref{trapro1}) the transformation law for
$f(x)$ and the form of $d(u)$ and $n(u)$
\begin{eqnarray}\label{ProTra}
  \tilde{f}(x) &=& a f(\Omega x + d_0) -2 (d_1^+\cdot\Omega x + n_0^+)
  \nonumber\\
  d(u)&=& d_0 + ud_1 + \theta(u) u d_1^+\nonumber\\
  n(u)&=& n_0 + \frac{u}{2} d_1^2 + \theta(u)(n_0^+ + \frac{u}{2}
  ({d_1^+}^2 + 2 d_1\cdot d_1^+)).
\end{eqnarray}
This extends the restricted nfp-subgroup (rnfp) found in \cite{AiBa2} 
of smooth nfp-transformations, preserving (\ref{DisPro}) by three additional 
parameters ($n_0^+,d_1^+$).

\section*{\Large\bit 2) Non-smooth rnfp-group and adjoint orbits}

The non-smooth rnfp transformations form a finite-dimensional
group whose elements $g$ are parametrized by
\begin{equation}\label{Para}
g=(a,\Omega;d_0,d_1,d_1^+;n_0,n_0^+),
\end{equation}
which in turn gives rise to the parameterization of the multiplication map 
$g_3=g_2 g_1$
\begin{eqnarray}\label{Mult}
  a_3&=&a_2 a_1\nonumber\\
  \Omega_3&=&\Omega_2\Omega_1\nonumber\\
  d_{03}&=&\Omega_2 d_{01} + d_{02}\nonumber\\
  d_{13}&=&\Omega_2 d_{11} + a_1 d_{12}\nonumber\\
  d_{13}^+&=&\Omega_2 d_{11}^+ + a_1 d_{12}^+\nonumber\\
  n_{03}&=&n_{01} + a_1 n_{02} + a_1\Omega_2 d_{01}\cdot d_{12}\nonumber\\
  n_{03}^+&=& n_{01}^+ + a_1 n_{02}^+ + a_1\Omega_2 d_{01}\cdot d_{12}^+.
\end{eqnarray}

\noindent
The second index $i=1,2,3$ in $d$ and $n$ refers to the group element $g_i$
to which it is associated.
The physical equivalence of profiles under rnfp-transformations is reflected 
in the equivalence of Killing vectors under the corresponding action of the 
rnfp-group. At the level of the group this amounts to the adjoint
action of the non-smooth rnfp-group on its Lie algebra.
>From (\ref{Mult}) it is easy to obtain the adjoint action of an arbitrary 
group element element $g$ on the Lie-algebra of the non-smooth rnfp-group, 
whose general element $X$ will be parametrized by
\begin{equation}\label{AlgPar} 
  X=(\alpha,\omega;D_0,D_1,D_1^+; N_0,N_0^+).
\end{equation}
\begin{eqnarray}\label{AdjAct}
  \tilde{\alpha}&=&\alpha\hspace{6cm}\tilde{X}=Ad(g)X\nonumber\\
  \tilde{\omega}&=&\omega\nonumber\\
  \tilde{D_0}&=&\Omega D_0 -\omega d_0\nonumber\\
  \tilde{D_1}&=&\frac{1}{a}(\Omega D_1 - \omega d_1 + \alpha d_1)\nonumber\\
  \tilde{D_1^+}&=&\frac{1}{a}(\Omega D_1^+ - \omega d_1^+ + 
  \alpha d_1^+)\\
  \tilde{N_0}&=&\frac{1}{a}(N_0 + \alpha n_0 + \Omega D_0\cdot d_1 - 
  d_0\cdot\Omega D_1 -\alpha d_0\cdot d_1 -\omega d_0\cdot d_1)\nonumber\\
  \tilde{N_0^+}&=&\frac{1}{a}(N_0^+ + \alpha n_0^+ + \Omega D_0\cdot d_1^+ - 
  d_0\cdot\Omega D_1^+ -\alpha d_0\cdot d_1^+ -\omega d_0\cdot d_1^+)\nonumber
\end{eqnarray}

The infinitesimal form of the transformation law of the profile 
is obtained from (\ref{ProTra}) by considering
$g$ to be an arbitrary curve through the unity with tangent $X$
\begin{equation}\label{VarPro}
  \delta f = \alpha f - x\omega\partial f + (D_0\partial) f 
  - 2(D_1^+ x + N_0^+).
\end{equation}
If in addition we require $X$ to be Killing the variation $\delta f$ of 
the profile has to vanish.
Since neither $D_1$ nor $N_0$ appear in  (\ref{VarPro}) they are immediately
recognized as Killing parameters thereby giving rise to a three-dimensional 
Killing algebra for the generic profile $f(x)$. Moreover, this 
three-dimensional subalgebra forms an abelian ideal within the Lie algebra of 
non-smooth rnfp transformations.  Consequently only the corresponding 
quotient algebra, whose general element may be characterized by setting 
$D_1$ and $N_0$ to 
zero in (\ref{AlgPar}), acts effectively on the space of profiles. 
Therefore the problem of characterizing the adjoint orbits of the full
group in the full algebra is reduced to the corresponding
subgroup and subalgebra respectively. 
Explicitly this procedure results in 
\begin{eqnarray}\label{AdAct}
  \tilde{\alpha}&=&\alpha\nonumber\\
  \tilde{\omega}&=&\omega\nonumber\\
  \tilde{D}_0&=&\Omega D_0 -\omega d_0\\
  \tilde{D}_1^+&=&\frac{1}{a}(\Omega D_1^+ -\omega d_1^+ + \alpha d_1^+)
  \nonumber\\
  \tilde{N}_0^+&=&\frac{1}{a}(N_0^+ + \alpha n_0^+ +\Omega D_0\cdot d_1^+
  -\Omega D_1^+\cdot d_0 - \alpha d_0\cdot d_1^+ - \omega d_0\cdot d_1^+),
  \nonumber
\end{eqnarray}
which gives rise to the following classification of adjoint orbits:

\noindent
\begin{itemize}
\item[1)] \underline{$\omega\neq 0,\alpha\neq 0$}
  \vspace{0.2cm}\\
  $X=(1,\omega_1;0,0,0;0,0)$
  \vspace{0.2cm}\\
  $Stab(X)=\{(a,\Omega;0,0,0;0,0)\}$
\item[2)] \underline{$\omega\neq 0,\alpha=0,N_0^+\neq 0$}
  \vspace{0.2cm}\\
  $X=(0,\omega_1;0,0,0;0,1)$
  \vspace{0.2cm}\\
  $Stab(X)=\{(1,\Omega;0,0,0;0,n_0^+)\}$
\item[3)] \underline{$\omega\neq 0,\alpha=0,N_0^+=0$}
  \vspace{0.2cm}\\
  $X=(0,\omega_1;0,0,0;0,0)$
  \vspace{0.2cm}\\
  $Stab(X)=\{(a,\Omega;0,0,0;0,n_0^+)\}$
\item[4)] \underline{$\omega=0,\alpha\neq 0, D_0\neq 0$}
  \vspace{0.2cm}\\
  $X=(1,0;D_0,0,0;0,0)$
  \vspace{0.2cm}\\
  $Stab(X)=\{(a,id;d_0,0,0;0,0)\}$
\item[5)] \underline{$\omega=0,\alpha\neq 0, D_0=0$}
  \vspace{0.2cm}\\
  $X=(1,0;0,0,0;0,0)$
  \vspace{0.2cm}\\
  $Stab(X)=\{(a,\Omega;d_0,0,0;0,0)\}$
\item[6)] \underline{$\omega=0,\alpha=0,D_0\neq 0,D_1^+\neq 0$}
  \vspace{0.2cm}\\
  $X=(0,0;D_0,0,D_1^+;0,0)$
  \vspace{0.2cm}\\
  $Stab(X)=\{(a,id;d_0,0,d_1^+;0,n_0^+)\}$ \qquad 
  with $D_0\cdot d_1^+=D_1^+\cdot d_0$
\item[7)] \underline{$\omega=0,\alpha=0,D_0\neq 0,D_1^+=0$}
  \vspace{0.2cm}\\
  $X=(0,0;D_0,0,0;0,0)$
  \vspace{0.2cm}\\
  $Stab(X)=\{(a,id;d_0,0,d_1^+;0,n_0^+)\}$ \qquad 
  with $D_0\cdot d_1^+=0$
\item[8)] \underline{$\omega=0,\alpha=0,D_0= 0,D_1^+\neq 0$}
  \vspace{0.2cm}\\
  $X=(0,0;0,D_1^+,0;0,0)$
  \vspace{0.2cm}\\
  $Stab(X)=\{(a,id;d_0,0,d_1^+;0,n_0^+)\}$ \qquad 
  with $D_1^+\cdot d_0=0$
\item[9)] \underline{$\omega=0,\alpha=0,D_0= 0,D_1^+=0,N_0^+\neq 0$}
  \vspace{0.2cm}\\
  $X=(0,0;0,0,0;0,1)$
  \vspace{0.2cm}\\
  $\label{tab}Stab(X)=\{(a,\Omega;d_0,0,d_1^+;0,n_0^+)\}$, 
\end{itemize}
\addtocounter{equation}{1}
where $X$ and $Stab(X)$ denote the canonical representative of the orbit  
and its stability group respectively.

\section*{\Large\bit 3) Profiles with four and more Killing vectors}

The task of finding profiles with a four-pa\-ra\-meter symmetry group
is now simply reduced to inserting the vectors $X$ into Killing's 
equation which determines the corresponding profile $f$. 

\begin{itemize}
\item[1)] $0=f - x\omega_1\partial f,\qquad f=\gamma_1\partial_\phi f$
  \vspace{0.2cm}\\
  \fbox{$f=h(\rho) e^{\frac{1}{\gamma_1}\phi}$}
\item[2)] $0= - x\omega_1\partial f -2 ,\qquad \gamma_1\partial_\phi f=-2$
  \vspace{0.2cm}\\
  \fbox{$f=h(\rho) -\frac{2}{\gamma_1}\phi$}
\item[3)] $0= - x\omega_1\partial f ,\qquad \partial_\phi f=0$
  \vspace{0.2cm}\\
  \fbox{$f=h(\rho)$}
\item[4)] $f +(D_0\partial) f=0$
  \vspace{0.2cm}\\
  \fbox{$f=h(\tilde{D}_0x)\exp(-\frac{D_0x}{D_0^2})$}
\item[5)] $f=0 \Longrightarrow \mbox{flat space}$
\item[6)] $(D_0\partial) f -2 D_1^+x=0$
  \vspace{0.2cm}\\
  \fbox{$f=h(\tilde{D}_0x) + \frac{1}{(D_0^2)^2}\left[ 
        (D_1^+D_0)(D_0x)^2 + 2(D_1^+\tilde{D}_0)
        (\tilde{D}_0x)(D_0x)\right]$}
\item[7)] $(D_0\partial) f=0$
  \vspace{0.2cm}\\
  \fbox{$f=h(\tilde{D}_0x)$}
\item[8)] $-2D_1^+x=0\Longrightarrow D_1^+=0\Longrightarrow$ no solution
\item[9)] $-2N_0^+=0\Longrightarrow N_0^+=0\Longrightarrow$ no solution.
\begin{equation}\label{FourKill}\end{equation}
\end{itemize}
In cases 1) and 2) $\rho$ and $\phi$ denote circular polar coordinates.
>From the above table we see that the relevant cases with four Killing vectors
are 1) to 4) and 6) and 7).

In order to obtain classes with even higher symmetries we have
to combine the above cases, by imposing additional Killing vectors.
This, however, cannot be done by simply taking 
$X$ from the classes 1) to 9), since they may not attain their ``canonical''
form in the same system of coordinates. 
Suppose we would like to find spacetimes with symmetries from orbit 1
and orbit 2. We may start with the canonical representative $X_1$ from 
orbit 1, but then we have to take the generic representative from orbit 2, 
i.e.~$g X_2 g^{-1}$. However, since the stability groups do not change the 
correponding canonical form, we may use, without loss in generality, 
$\hat{g}X_2\hat{g}^{-1}$, where $\hat{g}=g_1 g g_2$, with $g_1$ and $g_2$
belonging to the respective stability groups of $X_1$ and $X_2$. We then
impose $\hat{g}X_2\hat{g}^{-1}$ to be Killing thereby restricting the 
profile found for $X_1$ further. This will be done for all combinations.
The $\hat{g}$ obtained by combining elements from two orbits are: 
\vspace{1cm}

\begin{itemize}
\item[\fbox{1+2}] $g_1gg_2 = 
  (a_1a,\Omega_1\Omega\Omega_2;\Omega_1d_0,0,\Omega_1d_1^+;0,n_{02}^+ 
  + n_0^+)$
  \vspace{0.2cm}\\
  $\hat{g}=(1,id;d_0,0,d_1^+;0,0)$
\item[\fbox{1+3}] $g_1gg_3 = 
  (a_1aa_3,\Omega_1\Omega\Omega_3;\Omega_1 d_0,0,a_3\Omega_1 d_1^+;0,n_{03}^+ 
  + a_3 n_0^+)$ 
  \vspace{0.2cm}\\
  $\hat{g}=(1,id;d_0,0,d_1^+;0,0)$
\item[\fbox{1+4}] $g_1gg_4 = 
  (a_1aa_4,\Omega_1\Omega;\Omega_1(\Omega d_{04} +  d_0),0,
  a_4\Omega_1 d_1^+;0,a_4 n_0^+ + a_4 d_1^+\cdot\Omega d_{04} ) $ 
  \vspace{0.2cm}\\
  $\hat{g}=(1,id;0,0,\Omega^{-1}d_1^+;0,n_0^+,-d_0^2)$
\item[\fbox{1+6}] $g_1gg_6 = 
  (a_1aa_6,\Omega_1\Omega;\Omega_1(\Omega d_{06} + d_0),0,
  \Omega_1(\Omega d_{16}+a_6 d_1^+);0,n_{06}^+ + a_6 n_0^+ + $\\
  \hspace*{4cm}$a_6 d_1^+
  \cdot\Omega d_{06}) $ \qquad $D_{06}\cdot d_{16}^+ = D_{16}^+\cdot d_{06}$  
  \vspace{0.2cm}\\
  $\hat{g}=(1,id;0,0,d_{16}^+ + \Omega^{-1}d_1^+;0,0)$ with 
  $\tilde{D}_{06}(d_{16}^+ + \Omega^{-1}d_1^+)=0$\\
\item[\fbox{1+7}] $g_1gg_7 = 
  (a_1aa_7,\Omega_1\Omega;\Omega_1(\Omega d_{07} + d_0),0,
  \Omega_1(\Omega d_{17}+a_7 d_1^+);0,n_{07}^+ + a_7 n_0^+ + $\\
  \hspace*{4cm}$a_7  d_1^+
  \cdot\Omega d_{07}) $ \qquad $D_{07}\cdot d_{17}^+ = 0$  
  \vspace{0.2cm}\\
  $\hat{g}=(1,id;0,0,d_{17}^+ + \Omega^{-1}d_1^+;0,0)$ with 
  $\tilde{D}_{07}(d_{17}^+ + \Omega^{-1}d_1^+)=0$\\
\item[\fbox{2+3}] $g_2gg_3 = 
  (aa_3,\Omega_2\Omega\Omega_3;\Omega_2 d_0,0,
  a_3\Omega_2 d_1^+;0,n_{03}^+ + a_3 n_0^+ + a_3 a n_{02}^+)$ 
  \vspace{0.2cm}\\
  $\hat{g}=(1,id;d_0,0,\frac{1}{a}d_1^+;0,0)$
\item[\fbox{2+4}] $g_2gg_4 = 
  (aa_4,\Omega_2\Omega;\Omega_2(\Omega d_{04} + d_0),0,
  a_4\Omega_2 d_1^+;0,a_4 (n_0^+ + a n_{02}^+ ) + 
  a_4 \Omega d_{04}\cdot d_1^+ )$ 
  \vspace{0.2cm}\\
  $\hat{g}=(1,id;0,0,\frac{1}{a}\Omega^{-1}d_1^+;0,0)$
\item[\fbox{2+6}] $g_2gg_6 = 
  (aa_6,\Omega_2\Omega;\Omega_2\Omega d_{06} + \Omega_2 d_0,0,
  \Omega_2\Omega d_{16}^+ + a_6\Omega_2 d_1^+;0,n_{06}^+ +$\\
  \hspace*{4cm}$a_6 (n_0^+ + a n_{02}^+ ) +
  a_6 \Omega d_{06}\cdot d_1^+) $ \qquad $D_{06}\cdot d_{16}^+ = 
  D_{16}^+\cdot d_{06}$ \vspace{0.2cm}\\
  $\hat{g}=(1,id;0,0,d_{16}^+ +\frac{1}{a}\Omega^{-1}d_1^+;0,0)$
  with $\tilde{D}_{06}(d_{16}^+ + \frac{1}{a}\Omega^{-1}d_1^+)=0$
\item[\fbox{2+7}] $g_2gg_7 = 
  (aa_7,\Omega_2\Omega;\Omega_2\Omega d_{07} + \Omega_2 d_0,0,
  \Omega_2\Omega d_{17}^+ + a_7\Omega_2 d_1^+;0,$\\
  \hspace*{3cm}$n_{07}^+ + a_7 (n_0^+ + a n_{02}^+ ) + 
  a_7 \Omega d_{07}\cdot d_1^+ )$ \qquad $D_{07}\cdot d_{17}^+ = 0$
  \vspace{0.2cm}\\
  $\hat{g}=(1,id;0,0,d_{17}^+ +\frac{1}{a}\Omega^{-1}d_1^+;0,0)$
  with $\tilde{D}_{07}(d_{17}^+ + \frac{1}{a}\Omega^{-1}d_1^+)=0$
\item[\fbox{3+4}] $g_2gg_4 = 
  (a_3aa_4,\Omega_3\Omega;\Omega_3 d_0 + \Omega_3\Omega d_{04},0,
  a_4\Omega_3 d_1^+;0,a_4 (n_0^+ + a n_{03}^+ ))$ 
  \vspace{0.2cm}\\
  $\hat{g}=(1,id;0,0,\Omega^{-1}d_1^+;0,0)$
\item[\fbox{3+6}] $g_3gg_6 = 
  (a_3aa_6,\Omega_3\Omega;\Omega_3\Omega d_{06} + \Omega_3 d_0,0,
  \Omega_3\Omega d_{16}^+ + a_6\Omega_3 d_1^+;0,$\\
  \hspace*{2.5cm}$n_{06}^+ + a_6 (n_0^+ + a n_{03}^+ ) + 
  a_6 \Omega d_{06}\cdot d_1^+ $) \qquad $D_{06}\cdot d_{16}^+ = 
  D_{16}^+\cdot d_{06}$ \vspace{0.2cm}\\
  $\hat{g}=(1,id;0,0,d_{16}^+ +\Omega^{-1}d_1^+;0,0)$
  with $\tilde{D}_{06}(d_{16}^+ + \Omega^{-1}d_1^+)=0$
\item[\fbox{3+7}] $g_3gg_7 = 
  (a_3aa_7,\Omega_3\Omega;\Omega_3\Omega d_{07} + \Omega_3 d_0,0,
  \Omega_3\Omega d_{17}^+ + a_7\Omega_3 d_1^+;0,$\\
  \hspace*{3cm} $n_{07}^+ + a_7 (n_0^+ + a n_{03}^+ ) + 
  a_7 \Omega d_{07}\cdot d_1^+) $ \qquad $D_{07}\cdot d_{17}^+ = 0$
  \vspace{0.2cm}\\
  $\hat{g}=(1,id;0,0,d_{17}^+ +\Omega^{-1}d_1^+;0,0)$
  with $\tilde{D}_{07}(d_{17}^+ + \Omega^{-1}d_1^+)=0$
\item[\fbox{4+6}] $g_4gg_6 = 
  (a_4aa_6,\Omega;\Omega d_{06} + d_0 + d_{04},0,
  \Omega d_{16}^+ + a_6d_1^+;0,n_{06}^+ + a_6 n_0^+  + 
  a_6 \Omega d_{06}\cdot d_1^+) $ 
  \qquad $D_{06}\cdot d_{16}^+ = D_{16}^+\cdot d_{06}$ 
  \vspace{0.2cm}\\
  $\hat{g}=(1,\Omega;0,0,0;0,0)$
\item[\fbox{4+7}] $g_4gg_7 = 
  (a_4aa_7,\Omega;\Omega d_{07} + d_0 + d_{04},0,
  \Omega d_{17}^+ + a_7 d_1^+;0,n_{07}^+ + a_7 n_0^+  + 
  a_7 \Omega d_{07}\cdot d_1^+) $ \qquad $D_{07}\cdot d_{17}^+ = 0$
  \vspace{0.2cm}\\
  $\hat{g}=(1,id;0,0,\Omega(d_{17}^+ +\Omega^{-1}d_1^+);0,0)$
  with $\Omega^{-1}\tilde{D}_{07}(d_{17}^+ + \Omega^{-1}d_1^+)=0$
\item[\fbox{6+7}] $g_6gg_7 = 
  (a_6aa_7,\Omega;\Omega d_{07} + d_0 + d_{06},0,
  \Omega d_{17}^+ + a_7 d_1^+ + aa_7 d_{16}^+;0,$\\
  \hspace*{3.5cm} $n_{07}^+  +
  a_7(n_0^+ + a n_{06}^+) + a d_0\cdot d_{16}^+ a_7\Omega d_{07}\cdot 
  (d_1^+ a d_{16}^+)) $\\ 
  $D_{06}\cdot d_{16}^+ = D_{16}^+\cdot d_{06}; D_{07}\cdot 
  d_{17}^+ = 0$
  \vspace{0.2cm}\\
  $\hat{g}=(1,\Omega;0,0,0;0,0)$
\end{itemize}
Although this table seems rather long, the imposition of the 
$\hat{g}$-transformed Killing vector on the previously obtained
profiles (\ref{FourKill}) eliminates most of the above combinations.
The only non-trivial ones together with the corresponding 
relative transformations $\hat{g}$ are
\begin{itemize}
\item[\fbox{3+6}] 
  $f = h_0 + \lambda \rho^2$, \vspace{0.2cm} \\
  $X_3=(0,\omega_1;0,0,0;0,0)$\qquad $\hat{g}=id$,\\
  $X_6=(0,0;D_0,0,\lambda D_0;0,0)$,  
\item[\fbox{4+7}] 
  $f = h_0\exp{\frac{1}{D_0^2}\left( (\tilde{D}_0x)\frac{D_0\cdot\Omega D_{07}}
      {\tilde{D}_0\cdot\Omega D_{07}} - D_0x \right)}$, \vspace{0.2cm}\\
  $X_4=(1,0;D_0,0,0;0,0)$\qquad $\hat{g}=\Omega$,\\
  $X_7=(0,0;\Omega D_{07},0,0;0,0)$,
\item[\fbox{6+7}] 
  $f= h_0 + \frac{1}{(D_0^2)^2}\left[ \frac{(D_1^+\tilde{D}_0)^2}{D_1^+ D_0}
  (\tilde{D}_0x)^2 + (D_1^+ D_0)(D_0x)^2 + \right.$\\
  \hspace*{8.5cm}$\left. 2(D_1^+\tilde{D}_0)(D_0x)
  (\tilde{D}_0x)\right]$ \vspace{0.2cm},\\
  $X_6=(0,0;D_0,0,D_1^+;0,0)$\qquad $\hat{g}=\Omega$,\\
  $X_7=(0,0;\Omega D_{07},0,0;0,0)$.
\end{itemize}
Applying the same procedure to identical orbits, i.e.~imposing 
two vectors belonging to the same orbit, results in 6 combinations.
The non-trivial ones are given by
\begin{itemize}
\item[\fbox{$3+\bar{3}$}]$f = h_0 + \mu \rho^2$ \vspace{0.2cm} \\
  $X_3=(0,\bar{\omega}_1;0,0,0;0,0)$\qquad $\hat{g}=(1,id;d_0,0,\mu d_0;0,0)$\\
  ${X_{\bar{3}}}=(0,\bar{\omega}_1;-\bar{\omega}_1 d_0,0,-\mu\bar{\omega}_1
  d_0;0,0)$,
\item[\fbox{$4+\bar{4}$}]$f = h_0 \exp{ \frac{1}{D_0^2}
  \left[\frac{D_0(\Omega\bar{D}_0-D_0)}{\tilde{D}_0\Omega\bar{D}_0}\right ]} $
  \vspace{0.2cm} \\
  $X_4=(1,0;D_0,0,0;0,0)$\qquad $\hat{g}=\Omega$\\
  ${X_{\bar{4}}}=(1,0;\Omega \bar{D}_0,0;0,0)$,
\item[\fbox{$6+\bar{6}$}] 
  $f= h_0 + \frac{1}{(D_0^2)^2}\left[ 
  \frac{-(D_1^+\tilde{D}_0) + D_0^2(\tilde{D}_0\Omega\bar{D}_1^+)}
  {\Omega\bar{D}_0 \tilde{D}_0}
  (\tilde{D}_0x)^2 + (D_1^+ D_0)(D_0x)^2 +\right.$\\ 
  \hspace*{8.5cm}$\left. 2(D_1^+\tilde{D}_0)(D_0x)(\tilde{D}_0x)\right]$ 
  \vspace{0.2cm},\\
  $X_6=(0,0;D_0,0,D_1^+;0,0)$\qquad $\hat{g}=\Omega$\qquad 
        $D_1^+\cdot\Omega \bar{D}_0 = D_0\cdot \Omega\bar{D}^+_1$,\\
  $X_{\bar{6}} = (0,0;\Omega \bar{D}_0,0,\Omega \bar{D}_1^+;0,0)$.
\end{itemize}
This completes the classification for impulsive waves with a five-parametric 
Killing algebra.
It is however possible to recast them in a simpler form. 
\newline\noindent
\begin{minipage}[c]{6.5cm}
\begin{itemize}
\item[\fbox{3+6}] 
  $f = h_0 + \lambda \rho^2$, \vspace{0.2cm} \\
  $X_3=(0,\omega_1;0,0,0;0,0)$,\\
  $X_6=(0,0;D_0,0,\lambda D_0;0,0)$,  
\item[\fbox{4+7}] 
  $f = h_0\exp{\left(-\frac{\Omega\tilde{D}_{07}\cdot x}{D_{07}^2\lambda}
        \right)}$, 
\vspace{0.2cm}\\
  $\bar{X}_4=(1,0;\lambda \Omega \tilde{D}_{07},0,0;0,0)$\\
  $X_7=(0,0;\Omega D_{07},0,0;0,0)$,
\item[\fbox{6+7}] 
  $f= h_0 + \lambda\frac{(D_1^+ x)^2}{(D_1^+)^2}$
\vspace{0.2cm},\\
  $\bar{X}_6=(0,0;D_1^+,0,\lambda D_1^+;0,0)$\\
  $X_7=(0,0;\tilde{D}_1^+,0,0;0,0)$,
\end{itemize}
\end{minipage}\quad
\begin{minipage}[c]{7.5cm}
\begin{itemize}
\item[\fbox{3+$\bar{3}$}] 
  $f = h_0 + \mu \rho^2$, \vspace{0.2cm} \\
  $X_3=(0,\bar{\omega}_1;0,0,0;0,0)$,\\
  $X_6=(0,\bar{\omega}_1;E_0,0,\mu E_0;0,0)$\\
  $E_0=-\bar{\omega}_1d_0$,  
\item[\fbox{4+$\bar{4}$}] 
  $f = h_0\exp{\left(-\frac{\tilde{E}_0\cdot x}
  {D_0\tilde{E}_0} \right)}$, 
\vspace{0.2cm}\\
  $\bar{X}_4=(1,0;\frac{D_0\cdot E_0}{E_0^2}\tilde{E}_0,0,0;0,0)$\\
  $\bar{X}_{\bar{4}}=(0,0;E_0,0,0;0,0)$\\
  $E_0=\Omega\bar{D}_0 - D_0$,
\item[\fbox{$6+\bar{6}$}]
  $f(x) = h_0 + \\
\hspace*{1cm}\frac{1}{E_0^2}\left[ \lambda(E_0x)^2 +
\bar{\lambda}(\tilde{E}_0x)^2 \right]$\vspace{0.2cm}\\
  $\bar{X}_6 = (0,0;E_0,0,\lambda E_0;0,0)$\\
  $\bar{X}_{\bar{6}} = (0,0; \tilde{E}_0,0,\bar{\lambda} \tilde{E}_0;0,0)$.
\end{itemize}
\end{minipage}
\newline

\noindent
In order to obtain the above simplified form, we linearly combined those 
Killing vectors obtained directly from the classification process.
The simplification of $6+\bar{6}$ arises from noting that this class actually
represents a generic quadratic profile $f(x)=x\cdot Mx$, which
may be diagonalized giving rise to the eigenvalues $\lambda,\bar{\lambda}$.
It is interesting to note that the combination of elements belonging to
the same orbit is equivalent to the combination of different ones. 
So we see that there are actually only three different classes containing 
5 Killing vectors. Moreover, the Killing group of $3+\bar{3}$ is actually
six-dimensional due to the independence of the profile from $E_0$.\par
Proceeding along the same lines for the combination of three orbits, we 
find that there are no nontrivial cases.
Let us summarize the results in the following table\newline\noindent
\begin{tabular}{|l|l|l|l|} \hline
profile  &Killing vectors  &r &type \\ \hline\hline
$f=f(x)$ & $\xi_1=\partial_v$, 
        &3 &abelian \\ 
        &$\xi_2=2 x\partial_v + u\partial_x, $&&\\
        &$\xi_3=2y\partial_v + u\partial_y$&&\\ \hline
$f=h(\rho)e^{\frac{1}{\gamma_1}\phi}$ & $\xi_1,\xi_2,\xi_3,$&4
        &$[\xi_4,\xi_1] = \xi_1$,\\ 
        &$\xi_4=u\partial_u-v\partial_v-\gamma_1\partial_\phi$ &
        & $[\xi_4,\xi_2] = \xi_2 + \gamma_1 \xi_3$,\\ 
        &&&$[\xi_4,\xi_3] = \xi_3 - \gamma_1 \xi_2$\\\hline
$f=h(\rho)-\frac{2}{\gamma_1}\phi $ 
        &$\xi_1,\xi_2,\xi_3,$ &4& $[\xi_4,\xi_2] = \gamma_1\xi_3,$\\
        &$\xi_4 = -\gamma_1\partial_\phi + 2\theta(u)\partial_v$ 
        && $[\xi_4,\xi_3] = -\gamma_1\xi_2$ \\ \hline
$f=h(\rho)$ &$\xi_1,\xi_2,\xi_3,$  &4&$[\xi_4,\xi_2] = -\xi_3$,\\
        &$\xi_4=\partial_\phi$&&$[\xi_4,\xi_3] = \xi_2$\\ \hline
$f=h(y) e^{-\frac{1}{\lambda}x}$ &$\xi_1,\xi_2,\xi_3,$  &4 
        &$[\xi_4,\xi_1] = \xi_1$, \\
        &$\xi_4 = u\partial_u -v\partial_v + \lambda \partial_x  
        $&&$[\xi_4,\xi_2] = \xi_2 +2\lambda\xi_1$,\\
        &&&$[\xi_4,\xi_3] = \xi_3 $ \\ \hline
$f=h(y)+\left( \lambda x^2 + 2\bar{\lambda}xy \right) $&$\xi_1,\xi_2,\xi_3,$ 
        &4&$[\xi_4,\xi_2] = 2\xi_1$\\
        &$\xi_4 = \partial_x + \theta(u)(\lambda\xi_2 +
        \bar{\lambda}\xi_3 )$ &&\\ \hline
$f=h(y) $&$\xi_1,\xi_2,\xi_3,$ &4&$[\xi_4,\xi_2] = 2\xi_1$\\
        &$\xi_4 = \partial_x $&&\\ \hline
$f=h_0 + \lambda \rho^2$&$\xi_1,\xi_2,\xi_3,$&6
        &$[\xi_4,\xi_2] = 2\xi_1$,\\
        &$\xi_4=\partial_x + \lambda\theta(u)\xi_2,$&
        &$[\xi_4,\xi_6] =  \lambda\theta(u)\xi_3,$\\
        &$\xi_5=\partial_y + \lambda\theta(u)\xi_3$&
        &$[\xi_5,\xi_3] = 2\xi_1$,\\ 
        &$\xi_6=\partial_\phi$,&&$[\xi_5,\xi_6] = -\lambda\theta(u)\xi_2$\\ 
        &&&$[\xi_6,\xi_2] = -\xi_3$\\
        &&&$[\xi_6,\xi_3] = \xi_2$\\ \hline
$f=h_0 e^{-\frac{1}{\lambda}x}$&$\xi_1,\xi_2,\xi_3,$&5
        &$[\xi_4,\xi_1] = \xi_1$,\\
        &$\xi_4=u\partial_u-v\partial_v +\lambda \partial_x$,&
        &$[\xi_4,\xi_2] = \xi_2 + 2\lambda\xi_1$,\\ 
        &$\xi_5=\partial_y$
        &&$[\xi_4,\xi_3] = \xi_3$,\\ 
        &&&$[\xi_5,\xi_3] = 2\xi_1$\\ \hline
$f=h_0 + \left(\lambda x^2 +\bar{\lambda}y^2\right) $&$\xi_1,\xi_2,\xi_3,$&5
        &$[\xi_4,\xi_2] = 2\xi_1$\\
        &$\xi_4=\partial_x + \lambda\theta(u)\xi_2$,
        &&$[\xi_5,\xi_3] = 2\xi_1$\\
        &$\xi_5=\partial_y + \bar{\lambda}\theta(u)\xi_3$&&\\ \hline
\end{tabular}
\vspace{0.2cm}

\noindent
Let us remark that the above table lists only the non-vanishing commutators 
and that the profile is not required to obey any field-equations.
Specifically the vacuum equations require $f$ to be harmonic.
\section*{\Large\bit 4) Rosen form and differentiability conditions}

This section is devoted to the so-called Rosen form of the metric
(\ref{NF}). It amounts in the elimination of the $\delta$-term and 
brings the metric into a continuous form, however,
at the price of a discontinuous profile-dependent coordinate transformation.
At first sight this procedure might seem somewhat paradoxical
since the metric (\ref{NF}) is clearly discontinuous. However, one has to keep 
in mind that the discontinuous change of coordinates alters the differentiable
structure. With respect to the new differentiable structure, i.e.~forgetting
where the coordinates came from, the metric is $C^0$. 
The explicit form of the transformation is given by
\begin{eqnarray}\label{RoCo}
  u&=&U\nonumber,\\
  v&=&V+\theta(U)f(X) +U\theta(U)\frac{1}{4}(\partial f)^2,\nonumber\\
  x&=&X + U\theta(U)\frac{1}{2}\partial f.
\end{eqnarray}
Applying this transformation to (\ref{NF}) leaves us with
\begin{equation}\label{BF}
  ds^2= -dUdV + (\delta_{ij} + \frac{1}{2}U\theta(U)\partial_i\partial_jf)^2 
dX^i dX^j,
\end{equation}
which is obviously $C^0$. Since it is not possible
to explicitly invert (\ref{RoCo}) for a general profile $f$ 
let us illustrate the above remarks by considering an impulsive plane 
wave as a specific example.
Its profile is given by
\begin{equation}\label{Plane}
 f(x)=\frac{1}{E_0^2}\left( \lambda (E_0x)^2  + \bar{\lambda}(\tilde{E}_0x)^2 
 \right)
\end{equation}
Inserting this $f$ into (\ref{RoCo}) allows us to read off the inverse 
transformation:
\begin{eqnarray}\label{Inbrco}
  U&=&u\nonumber,\\
  V&=&v-\frac{1}{E_0^2}\theta(u)\left(\frac{\lambda}{1+\lambda u} (E_0 x)^2
    + \frac{\bar{\lambda}}{1+\bar{\lambda} u} (\tilde{E}_0 x)^2\right ),
  \nonumber\\
  X&=&x - \frac{1}{E_0^2} u\theta(u)\left( 
    \frac{\lambda}{1+\lambda u} (E_0 x) E_0+ 
    \frac{\bar{\lambda}}{1+\bar{\lambda} u} (\tilde{E}_0 x)
    \tilde{E}_0 \right).
\end{eqnarray} 
However, the three Killing vectors that were smooth with respect to the 
original differentiable structure are now non-smooth.
\begin{eqnarray}\label{Coki}
\xi_1 &=& \partial_V \nonumber\\
\xi_2 &=& 2X\partial_V + U \partial_X
    - U^2\theta(U)\frac{\lambda}{1+\lambda U}\partial_X\nonumber\\
\xi_3 &=& 2Y\partial_V + U \partial_Y
    - U^2\theta(U)\frac{\bar{\lambda}}{1+\bar{\lambda} U}\partial_Y
\end{eqnarray}
Whereas those that were discontinuous now become smooth.
\begin{eqnarray}\label{discoki}
  \xi_4 &=& (1+\lambda u\theta(u)) \partial_x + 2\lambda \theta(u)
  x\partial_v = \partial_X\nonumber\\
  \xi_5 &=& (1+\bar{\lambda} u\theta(u)) \partial_y + 
  2\bar{\lambda} \theta(u) y\partial_v = \partial_Y
\end{eqnarray}

\newpage
\noindent
{\Large\bit Conclusion}\newline

In this work we analyzed the symmetry structure of pp-waves with
an  impulsive profile (i.e.~waves where the curvature is concentrated on a
null surface). The mathematical framework describing such an idealized 
physical situation is necessarily distributional 
in nature. Therefore an investigation of possible Killing vectors is 
immediately confronted with the question about the domain of definition of 
the Killing equation. In general this equation contains products
of the Killing vector and the metric which restricts the Killing vector
to be $C^\infty$. However, due to the specific form of the metric 
under consideration, it was possible to extend the class of Killing vectors 
to contain also non-smooth fields while at the same time maintaining a 
distributionally well-defined structure. This extension lead us to a 
larger class of symmetries which on physical grounds we think is necessary 
to include.
Consider for instance the plane wave (the last case in our table), which
has only three smooth Killing vectors, namely $\xi_1, \xi_2, \xi_3$, 
while the corresponding non-impulsive wave in the JEK-classification admits a 
five-parametric isometry group. In the limit to an impulsive wave one 
exactly recovers the two ``missing'' Killing vectors $\xi_4$ and $\xi_5$ as 
non-smooth symmetries.
On the other hand it is in general not possible to  obtain the
isometry group by taking the limit from a pp-wave with a 
non-impulsive profile. This is precisely the reason why the JEK-classification
fails and additional symmetries show up.
One may however ask if those spacetimes which cannot be 
obtained from boosting a sandwich-wave are physically meaningful.
We would like to give an affirmative answer, e.g.~the second case in our 
table, as pointed out in \cite{AiBa2} becomes in the vacuum case the 
so-called AS-metric, whose symmetry-group can be obtained by boosting the 
Killing vectors of the Schwarzschild-geometry to the velocity of light 
\cite{AiBa1}.

Finally we addressed the question of the so-called 
Rosen-coordinates with respect to which the metric becomes $C^0$.
The corresponding change of coordinates is of course discontinuous and
turns Killing vectors that were smooth with respect to the original
differentiable structure into non-smooth vector fields with respect
to the new structure. \par
A natural extension of our work would be the investigation
the geodesic-equation for impulsive pp-waves, which faces the
same definition problems. It seems, however, that 
the correct setting for this problem requires more sophisticated 
techniques \cite{Col}, which allow to control products of
distributions. Work in this direction is currently in progress.  

\end{document}